\documentclass{IEEEtran}
\usepackage{graphicx}
\usepackage{amsmath}
\usepackage{amsfonts}
\usepackage{color}

\newcommand{\vc}[1]{{\boldsymbol{#1}}}
\newcommand{\mat}[1]{{\boldsymbol{#1}}}
\newcommand{\G}{{\mathcal G}}
\newcommand{\dd}{{\mathrm d}}
\newcommand{\cc}{{\mathrm c}}
\newcommand{\ee}{{\mathrm e}}

\newcommand{\Bd}{{\mathbb B}}
\newcommand{\Td}{E_\dd}

\newcommand{\sinc}{\mathrm{sinc}}
\newcommand{\samp}[1]{{\mathcal S}_{#1}}
\newcommand{\lift}{\mathcal L}
\newcommand{\ilift}{{\mathcal L}^{-1}}
\newcommand{\Real}{\mathbb{R}}
\newcommand{\Z}{\mathbb{Z}}
\newcommand{\Kid}{K_{\mathrm{id}}}
\newcommand{\Kidtf}{\hat{K}_{\mathrm{id}}}
\newcommand{\kid}{k_{\mathrm{id}}}
\newcommand{\Nyquist}{\mathrm{N}}
\newcommand{\jj}{\mathrm{j}}

\newcommand{\ind}{{\boldsymbol{1}}}

\newcommand{\tw}{\widetilde{w}}
\newcommand{\tv}{\widetilde{v}}
\newcommand{\E}{{E}}
\newcommand{\tE}{\mathcal{E}}
\newcommand{\tl}{{\tilde{\ell}}^2}
\newcommand{\B}{{\mathcal B}}

\newtheorem{deff}{Definition}
\newtheorem{theorem}{Theorem}
\newtheorem{lemma}{Lemma}
\newtheorem{problem}{Problem}
\newtheorem{remark}{Remark}
\newtheorem{prop}{Proposition}

\begin{document}
\title{$H^\infty$-Optimal Fractional Delay Filters}
\author{
\thanks{
Copyright (c) 2012 IEEE. Personal use of this material is permitted. 
However, permission to use this material 
for any other purposes must be obtained 
from the IEEE by sending a request to pubs-permissions@ieee.org.
}
Masaaki~Nagahara\thanks{M.~Nagahara is with 
Department of Applied Analysis and Complex Dynamical Systems,
Graduate School of Informatics, Kyoto University, Kyoto, 606-8501, JAPAN 
    (e-mail: {\tt nagahara@ieee.org}).
    Mailing address: Kyoto University, Yoshida Honmachi, 
    Sakyo-ku, Kyoto 606-8501, Japan.  
    },~\IEEEmembership{Member,~IEEE},
Yutaka~Yamamoto\thanks{Y.~Yamamoto is with 
Department of Applied Analysis and Complex Dynamical Systems, 
Graduate School of Informatics, Kyoto University, Kyoto, 606-8501, JAPAN 
    (e-mail: {\tt yy@i.kyoto-u.ac.jp}).
    Mailing address: Kyoto University, Yoshida Honmachi, 
    Sakyo-ku, Kyoto 606-8501, Japan.  
    },~\IEEEmembership{Fellow,~IEEE}%
}%
\markboth{}{}
\maketitle

\begin{abstract}
Fractional delay filters are digital filters to delay
discrete-time signals by a fraction of the sampling period.  
Since the delay is fractional, the intersample behavior 
of the original analog signal becomes crucial.  
In contrast to the conventional designs based 
on {the} Shannon sampling theorem with the band-limiting {hypothesis},
the present paper proposes a new approach 
based on the modern sampled-data $H^{\infty}$ optimization 
that aims at restoring the intersample behavior
beyond the Nyquist frequency.  
By using {the} lifting transform or continuous-time blocking
the design problem is equivalently reduced to a 
discrete-time $H^\infty$ optimization,
which can be effectively solved by numerical computation softwares.
Moreover, a closed-form solution is obtained 
under an assumption on the original analog signals.
Design examples are given to illustrate the advantage of the proposed method.
\end{abstract}

\IEEEpeerreviewmaketitle

\section{Introduction}
\label{sec:introduction}
Fractional
delay filters are digital filters to
delay discrete-time signals by a fractional amount of the sampling
period.
Such filters have wide applications in signal processing, including
sampling rate conversion~\cite{Ram84,SmiGos84,ParHilRob91},
nonuniform sampling~\cite{JohLow02,PreLevHur04},
wavelet transform~\cite{Yu07,Dum08},
digital modeling of musical instruments~\cite{ValLehLaa01,Val+06},
to name a few.
For more applications, see survey papers~\cite{LVKL96,ValLaa00,ValLaa07}.

Conventionally, fractional delay filters 
are designed
based on {the} Shannon sampling theorem~\cite{Sha49,Uns00}
for strictly bandlimited analog signals.
Based on this theory, the optimal filter coefficients
are {obtained} by sampling a delayed sinc function.
This ideal low-pass filter is however not realizable
because of its non-causality and instability,
and hence many studies have focused their attention on 
approximating the ideal filter by, for example, 
windowed sinc functions~\cite{CaiYarHen95,Sel08},
maximally-flat 
FIR approximation~\cite{Her92,PeiWan01,SamAhmSwa04,HacGunKon07,ShyPei08},
all-pass approximation~\cite{Jin87,PeiWan04},
and minmax (Chebyshev) optimization~\cite{PutSmi97}.

In particular, 
$H^2$ (or weighted least-squares) design has been prevalent in the 
literature~\cite{LVKL96,TarCaiHerRoj97,DenNak04,ShyPei08}.  
This method minimizes the $H^2$ norm of the weighted
difference between the ideal low-pass filter and a filter to be designed, 
and is based on the projection theorem in Hilbert space.  
There are, however, two major drawbacks in this 
conventional approach.  One is that due to the averaging
nature of the $H^{2}$ criterion, the obtained frequency 
response can have a sharp peak at a certain frequency, 
thereby yielding a poor performance at that frequency, 
while still maintaining small $H^{2}$ error in the overall frequency 
response.  The other is that $H^{2}$ criterion can yield 
a truncated frequency response as an optimal 
approximant of the ideal low-pass filter, which yields a 
distortion due to the Gibbs phenomenon in the time domain.
Furthermore, such a design 
is mostly executed in the discrete-time domain, which yields 
poor intersample response.  

In view of these problems we employ 
{\em sampled-data $H^\infty$ optimization,} 
recently introduced for signal processing 
by~\cite{YamNagKha12}%
\footnote{The approach dates back to~\cite{PPKYYCDC95},
though.}.  This is based on 
{\em sampled-data control theory}~\cite{CheFra} 
which accounts for the mixed nature of 
{\em continuous- and discrete-time\/} thereby enabling 
optimization of the intersample signals via discrete-time
controllers (filters).  This also allows for optimization 
according to the $H^\infty$ norm, namely {\em minimizing the 
maximum of the error frequency response}.  
This worst-case design is clearly desirable in that 
it does not have the drawback due to the averaging property
of the $H^{2}$ criterion.  Due to the nature of the 
$H^{\infty}$ norm, however, this optimization problem 
has been difficult to solve, but one can now utilize 
a standardized method to solve this class of problems~\cite{DGKF,CheFra}.
Furthermore, the obtained 
filter shows greater robustness against unknown 
disturbances due to the nature of the {\em uniform 
attenuation\/} of the error frequency response; 
see~\cite{YamNagKha12} for details.  
Based on this $H^{\infty}$ optimization method, 
we formulate the design of fractional delay filters
as a sampled-data $H^\infty$ optimization problem%
\footnote{
 This method was first proposed in our conference articles~\cite{NagYam03,NagYam05}.
 The present paper reorganizes these works
 with new results on the state-space formulation
 (Proposition \ref{prop:ss}, Appendix \ref{app:proof1}).
 Simulation results in Section \ref{sec:example} are also new.
}.

In order to optimize the intersample behavior, 
we must deal with both continuous- and discrete-time signals, 
and hence the overall system is not time-invariant.
The key to solving this problem is {\em lifting},
which is introduced in the {early} studies of {modern} 
sampled-data control theory~\cite{Yam90,BamPeaFraTan91,Toi92,BamPea92,Yam94}.
Indeed, continuous-time lifting gives an exact, not approximated{,}
time-invariant discrete-time {model} 
for a sampled-data {system}, albeit with infinite-dimensional
input and output spaces.  
Hence the problem of the mixed 
time sets is circumvented without approximation.  

Lifting can also be interpreted as a 
{\em continuous-time blocking or polyphase decomposition}.  
As in multirate signal processing~\cite{Vai},
lifting makes it possible to capture 
continuous-time signals and systems 
in the discrete-time domain without approximation; 
see Section \ref{subsec:lifting} for details.
The remaining system becomes a {\em time-invariant discrete-time\/} system,
albeit with infinite-dimensional input and output spaces.  
In view of this infinite-dimensionality, we retain the
term lifting to avoid confusion.  
With such a representation, we show that 
our design problem is reducible 
to a finite-dimensional discrete-time $H^\infty$ optimization 
without approximation.  
This type of $H^\infty$ optimization is easily {solvable} 
by standard softwares such as MATLAB~\cite{RCToolbox}.  

In some applications, digital filters with variable delay responses
({\em variable fractional delay filters}~\cite{Far88,LVKL96,TarCaiHerRoj97,DenNak04})
are desired.
{In this case, a filter should have a tunable}
{delay parameter, and hence a closed-form formula should}
{be derived.
In general,
$H^\infty$ optimal filters are difficult to}
{solve analytically.
However, we provide a closed-form}
{formula of the optimal filter
with the delay variable as a}
{parameter under the assumption that the underlying }
{frequency characteristic of continuous-time input signals}
{is governed by a low-pass filter of first order.}
While this assumption may appear somewhat restrictive, 
it covers many typical cases and variations 
by some robustness properties
\cite{NagYam03}.

The paper is organized as follows.
Section \ref{sec:problem} defines fractional delay filters, and
reviews a standard $H^{2}$ design method.  
We then reformulate our design problem as a sampled-data $H^\infty$ 
optimization to overcome the difficulty due to the $H^{2}$ design.  
Section \ref{sec:design} gives a procedure to solve the
sampled-data $H^\infty$ optimization problem
based on the lifting transform.
Section \ref{sec:example} shows numerical examples
to illustrate the superiority of the proposed method.

\subsection*{Notation}
Throughout this paper, we use the following notation.
We denote by $L^2[0,\infty)$ and $L^2[0,T)$ the Lebesgue spaces 
consisting of all square integrable real functions
on $\Real_{+}:=[0,\infty)$ and $[0,T)$, respectively.
$L^2[0,\infty)$ may be abbreviated {as} $L^2$.
By $\ell^2$ we denote the set of all real-valued square summable sequences 
on $\Z_{+}:=\{0,1,2,\ldots\}$.
For a normed space $X$, we denote by $\ell^2(\Z_{+},X)$ the set of all
sequences on $\Z_{+}$ taking values 
in $X$ with squared norms being summable.  
For {normed} linear spaces $X$ and $Y$,
we denote by $\Bd(X,Y)$ the set of all bounded 
linear operators of $X$ into $Y$.
$\Real^\nu$ and $\Real^{m\times n}$
denote respectively the set of real vectors
of size $\nu$ and real matrices of size $m\times n$.
Finite-dimensional vectors and matrices are denoted by bold letters,
such as $\vc{x}$ or $\vc{A}$, and
infinite-dimensional operators by calligraphic letters, 
such as ${\mathcal{B}}$.
The transpose of a matrix $\mat{A}$ is denoted by $\mat{A}^\top$.
Symbols $s$ and $z$ are used for the variables of 
Laplace and $Z$ transforms, respectively.
For a linear system $F$, its transfer function is denoted by
$\hat{F}(z)$ (if $F$ is discrete-time) or 
$\hat{F}(s)$ (if $F$ is continuous-time),
and its impulse response by the lower-case letter, $f[n]$ or $f(t)$.
The imaginary unit $\sqrt{-1}$ is denoted by $\jj$.

\section{Fractional Delay Filters}
\label{sec:problem}

In this section, we review fractional delay filters
with conventional design methods based on
{the} Shannon sampling theorem.
Then, we reformulate the design problem as
{a} sampled-data $H^\infty$ optimization {problem}.

\subsection{Definition and standard design method}

\begin{figure}[t]
 \centering
 \includegraphics[width=\linewidth]{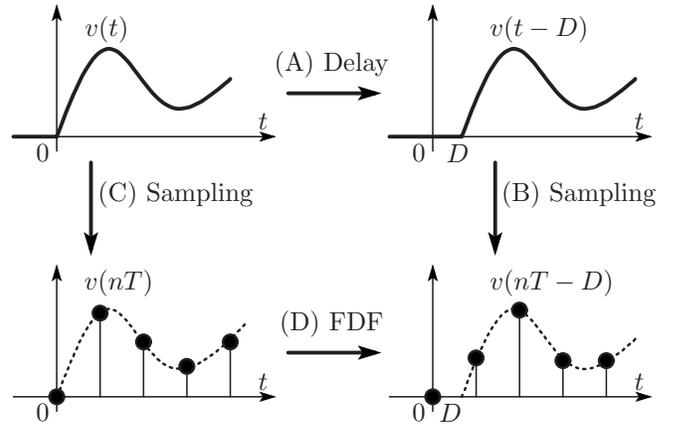}
 \caption{Fractional delay process:
  (A) a continuous-time signal $v(t)$ (top left) is delayed by $D>0$.
  (B) the delayed signal $v(t-D)$ is sampled at $t=nT$, $n=0,1,\ldots$.
  (C) the signal $v(t)$ is sampled at $t=nT$, $n=0,1,\ldots$.
  (D) digital filtering (fractional delay filter, FDF)
  to produce (or estimate) the sequence $\{v(nT-D)\}$
  from the sampled-data $\{v(nT)\}$.
 }
 \label{fig:fdelay}
\end{figure}

Consider a continuous-time signal $v$ shown 
in Fig.~\ref{fig:fdelay} (top-left figure).
Assume $v(t)=0$ {for} $t<0$ 
(i.e., {it is} a causal signal).
Delaying this signal by $D>0$ gives
the delayed continuous-time signal $v(t-D)$ 
shown in Fig.~\ref{fig:fdelay} (top-right {in Fig.~\ref{fig:fdelay}}).
Then by sampling $v(t-D)$ with sampling period $T$, 
we obtain {the} discrete-time signal $\{v(nT-D)\}_{n\in\Z}$
as shown in Fig.~\ref{fig:fdelay} (bottom-right {in Fig.~\ref{fig:fdelay}}).

Next, let us consider the sampled signal $\{v(nT)\}_{n\in\Z}$
of the original analog signal $v$
as shown in Fig.~\ref{fig:fdelay} (bottom-left {in Fig.~\ref{fig:fdelay}}).
The {objective} of fractional delay filters is
to reconstruct or estimate the delayed sampled signal
$\{v(nT-D)\}_{n\in\Z}$ {directly} from the sampled data $\{v(nT)\}_{n\in\Z}$
when $D$ is not an integer multiple of $T$.
We now define the {\em ideal} fractional delay filter.
\begin{deff}
 The {\em ideal fractional delay filter} $\Kid$ 
 with delay $D>0$ is the mapping that produces $\{v(nT-D)\}_{n\in\Z}$
 from $\{v(nT)\}_{n\in\Z}$,
 that is,
 \[
  \Kid: \{v(nT)\}_{n\in\Z}\mapsto \{v(nT-D)\}_{n\in\Z}.
 \]
\end{deff}

Assume {for the moment} that the original analog signal $v$ 
is fully band-limited 
below the Nyquist frequency $\Omega_{\Nyquist}=\pi/T$, 
that is,
\begin{equation}
 \hat{v}(\jj\omega) = 0, \quad |\omega|\geq \Omega_{\Nyquist},
 \label{eq:Nyquist}
\end{equation}
where $\hat{v}$ is the Fourier transform of $v$.
Then the impulse response of the ideal fractional delay filter is obtained 
by~\cite{LVKL96}:
\begin{equation}
 \begin{split}
  \kid [n]&=\frac{\sin \pi(n-D/T)}{\pi (n-D/T)} = \sinc(n-D/T),\\
  n&=0,\pm 1, \pm 2,\ldots,\quad \sinc(t) := \frac{\sin(\pi t)}{\pi t}.
 \end{split}
 \label{eq:ideal_FD}
\end{equation}
The frequency response of this ideal filter is given in the 
frequency domain as 
\begin{equation}
 \Kidtf(\ee^{\jj\omega T}) = \ee^{-\jj\omega D},\quad 
  \omega \leq \Omega_{\Nyquist}.
 \label{eq:ideal_FD_fourier}
\end{equation}

Since the impulse response \eqref{eq:ideal_FD}
does not vanish at $n=-1,-2,\ldots$ and is not absolutely summable,
the ideal filter {is} noncausal and unstable, and hence
the ideal filter {is not physically realizable}. 
Conventional designs thus aim at approximating
the impulse response \eqref{eq:ideal_FD} or 
the frequency response \eqref{eq:ideal_FD_fourier}
by a causal and stable filter.
We here review in particular the $H^2$ optimization,
also known as weighted least squares~\cite{LVKL96}.  

Define the weighted approximation error by 
\begin{equation}
 E_2:=(\Kid-K)W_\dd
 \label{eq:E2}
\end{equation}
where $W_\dd$ is a weighting function and $K$ is a filter to be designed, 
which is assumed to be FIR (finite impulse response).
The $H^2$ design aims at finding the FIR coefficients of the transfer function $\hat{K}(z)$ of $K$
that minimize the $H^2$ norm of the weighted error system $E_2$:
\begin{equation}
 \begin{split}
  \|E_2\|_2^2 
   &= \|(\Kid-K)W_\dd\|_2^2\\
   &= \frac{1}{\Omega_{\Nyquist}}
       \int_0^{\Omega_{\Nyquist}}
       \left|\left[
        \Kidtf(\ee^{\jj\omega T})-\hat{K}(\ee^{\jj\omega T})
       \right]
       \hat{W}_\dd(\ee^{\jj\omega T})\right|^2 \dd\omega.
 \end{split}
 \label{eq:H2norm}
\end{equation}

As pointed out in the Introduction, 
this $H^2$ design has some drawbacks.
One is that the designed filter $K$ may yield 
a large peak in the error frequency response 
$\hat{E}_2(\ee^{\jj\omega T})$ due to the 
averaging nature of the $H^{2}$ norm \eqref{eq:H2norm}.  
If an input signal has 
a frequency component at around such a 
peak of $\hat{E}_2(\ee^{\jj\omega T})$, the error will become 
very large.  
The second is that the perfect band-limiting assumption 
\eqref{eq:Nyquist} implies that the $H^{2}$ suboptimal filter 
is given as an approximant of the ideal low-pass filter~\cite{Fliege}, 
which induces large errors in the time domain~\cite{YamNagKha12}.  
Moreover, real analog signals always contain frequency 
components beyond the Nyquist frequency, 
and hence \eqref{eq:Nyquist} never holds exactly for real signals.  

\subsection{Reformulation of design problem}

To simultaneously solve the two problems pointed out above,
we introduce sampled-data $H^\infty$ optimization~\cite{YamNagKha12}.
This method has advantages as mentioned in Section \ref{sec:introduction}.
To adapt sampled-data $H^\infty$ optimization for 
the design of fractional delay filters,
we reformulate the design problem,
instead of mimicking the ``ideal'' filter given in \eqref{eq:ideal_FD}
or \eqref{eq:ideal_FD_fourier}.

\begin{figure}[t]
 \centering
 \includegraphics[width=\linewidth]{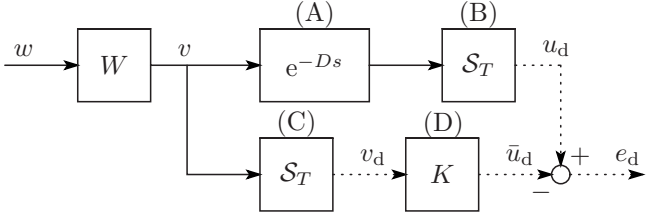}
 \caption{Error system $\E$ for designing fractional delay filter $K$. 
  (A)--(D) correspond to those in Fig.~\ref{fig:fdelay}.}
 \label{fig:error_sys}
\end{figure}

Let us consider the error system shown in Fig.~\ref{fig:error_sys}.
$W$ is a stable continuous-time system with 
strictly proper transfer function $\hat{W}(s)$
that defines the frequency-domain characteristic
of the original analog signal $v$.
More precisely, we assume that the analog signal $v$ 
is in the following subspace of $L^2$:
\[
 WL^2 := \left\{v\in L^2: v=Ww,~w\in L^2\right\}.
\]
Note that the signal subspace $WL^2$ is much wider
than that of band-limited $L^2$ signals~\cite{NagOguYam11}.

The upper path of the diagram in Fig.~\ref{fig:error_sys}
is the ideal process of 
the fractional delay filter 
(the process (A) $\rightarrow$ (B) in Fig.~\ref{fig:fdelay}); 
that is, the continuous-time signal $v$ is delayed 
by the continuous-time delay denoted by $\ee^{-Ds}$
(we use the notation $\ee^{-Ds}$, 
the transfer function of the $D$-delay system, as the system itself),
and then sampled 
by the ideal sampler denoted by $\samp{T}$ with period $T>0$
to become an $\ell^2$ signal%
\footnote{
 If $\hat{W}(s)$ is stable and strictly proper,
 the discrete-time signal $u_\dd=\samp{T}\ee^{-Ds}v$ belongs to $\ell^2$.
 Otherwise, $\samp{T}$ is not a bounded operator on $L^2$;
 see~\cite[Section 9.3]{CheFra}.
}
$u_\dd:=\samp{T}\ee^{-Ds}v$, or
\[
 u_\dd[n]:=\left(\samp{T}\ee^{-Ds}v\right)[n] = v(nT-D),\quad n\in\Z_+.
\]

On the other hand, the lower path {represents} the real process
((C) $\rightarrow$ (D) in Fig.~\ref{fig:fdelay}); 
that is, the continuous-time signal $v$ is 
directly sampled with the same period $T$
to produce a discrete-time signal $v_\dd\in\ell^2$ defined by
\[
 v_\dd[n]:=\left(\samp{T}v\right)[n]=v(nT),\quad n\in\Z_+.
\]
This signal is then filtered by a digital filter $K$ 
to be designed, and we obtain an estimation
signal
$\bar{u}_\dd = K\samp{T}v \in \ell^2$.

Put $e_\dd:=u_\dd-\bar{u}_\dd$ 
(the difference between the ideal output $u_\dd$ and 
the estimation $\bar{u}_\dd$), and
let $\E$ denote the error system from $w\in L^2$ to $e_\dd\in \ell^2$
(see Fig.~\ref{fig:error_sys}).
Symbolically, $\E$ is represented by 
(cf. \eqref{eq:E2})
\begin{equation}
 \E = \left(\samp{T}\ee^{-Ds}-K\samp{T}\right)W.
 \label{eq:E}
\end{equation}
Then our problem is to find a digital filter $K$ {that} minimizes 
the $H^\infty$ norm of the error system $\E$.
\begin{problem}
\label{prob:design}
 Given a stable, strictly proper $W(s)$, a delay time $D>0$,
 and a sampling period $T>0$,
 find the digital filter $K$ that minimizes 
(cf. \eqref{eq:H2norm})
 \[
  \begin{split}
   \|\E\|_\infty 
    &= \left\|\left(\samp{T}\ee^{-Ds}-K\samp{T}\right)W\right\|_\infty\\
    &= \sup_{w\in L^2,~\|w\|_2=1}
     \|\left(\samp{T}\ee^{-Ds}-K\samp{T}\right)W w\|_{\ell^2}.
  \end{split}
 \]
\end{problem}

Note that $W$, or its transfer function $\hat{W}(s)$,
can be interpreted
as a frequency-domain weighting function for the optimization.
This is comparable to $\hat{W}_\dd(z)$ in the discrete-time $H^2$ design
minimizing \eqref{eq:H2norm}.
The point to use continuous-time $\hat{W}(s)$ is that
one can model the frequency characteristic of signals
beyond the Nyquist frequency.  
Also, the advantage of using
the sampled-data setup here is that we can minimize the 
norm of the overall transfer operator from continuous-time $w$ 
to the error $e_{\rm d}$.  
In the next section, we will show a procedure to solve Problem 
\ref{prob:design}
based on sampled-data control theory.

\section{$H^\infty$ Design of Fractional Delay Filters}
\label{sec:design}

The error system $\E$ in Fig.~\ref{fig:error_sys}
contains both continuous- and discrete-time signals,
and hence the system is not time-invariant; 
in fact, it is $T$-periodic~\cite{CheFra}. 
In this section, 
we introduce the continuous-time lifting technique~\cite{Yam94,CheFra}
to derive a norm-preserving transformation
from $\E$ to a time-invariant finite-dimensional discrete-time system.
After this, one can use a standard discrete-time $H^\infty$ optimization
implemented on a computer software such as MATLAB
to obtain an optimal filter.
We also give a closed-form solution of the optimization under 
an assumption on $W(s)$.

\subsection{Lifted model of sampled-data error system}
\label{subsec:lifting}

\begin{figure}[t]
 \centering
 \includegraphics[width=\linewidth]{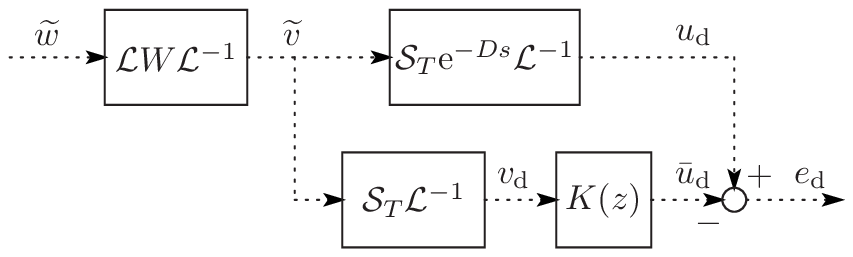}
 \caption{Lifted error system $\tE$.}
 \label{fig:lifted_sys}
\end{figure}

Let $\{\mat{A},\mat{B},\mat{C}\}$ be a minimal realization~\cite{Rug}
of {$\hat{W}(s)$:}
\begin{equation}
 \frac{\dd \vc{x}(t)}{\dd t} = \mat{A}\vc{x}(t) + \mat{B}w(t),~
 v(t) = \mat{C}\vc{x}(t),~t\in\Real_+,
 \label{eq:Fss}
\end{equation}
where $\vc{x}(t)\in\Real^\nu$ is the state variable 
($\nu$ is a positive integer).
We assume $\mat{A}\in\Real^{\nu\times\nu}$, 
$\mat{B}\in\Real^{\nu\times 1}$, 
$\mat{C}\in\Real^{1\times\nu}$, and 
$\vc{x}(0)=\vc{0}$.
Let $D=mT+d$ where $m\in\Z_{+}$ and $d$ is a real number
such that $0\leq d<T$.
First, we introduce the lifting operator $\lift$~\cite{Yam94,CheFra}
that transforms a continuous-time signal in $L^2[0,\infty)$ 
to an $\ell^2$ sequence 
of functions in $L^2[0,T)$.
Apply $\lift$ to the continuous-time signals 
$w$ and $v$, and put $\tw:=\lift w$, $\tv :=\lift v$. 
By this, 
the error system in Fig.~\ref{fig:error_sys},
is transformed into a time-invariant discrete-time system $\tE$
shown in Fig.~\ref{fig:lifted_sys}.
Since the operator $\lift$ gives {an isometry} 
between $L^2[0,\infty)$ and $\tl:=\ell^2(\Z_+,L^2[0,T))$, we have
\begin{equation}
\|\E\|_\infty 
  = \|\tE\|_\infty 
  := \sup_{\tw\in\tl,~\|\tw\|_\tl=1}
    \|\tE\tw\|_{\ell^2}
 \label{eq:EtotE}
\end{equation}

The following proposition is fundamental to 
the sampled-data optimization in \eqref{eq:EtotE}.
\begin{prop}
\label{prop:ss}
 A state-space realization of the lifted error system 
 $\tE$ is given by
 \begin{equation}
  \begin{split}
   \vc{\xi}[n+1] &= \mat{A}_\dd \vc{\xi}[n] 
	+ \left[\begin{array}{c}\B\\0\end{array}\right] \tw[n],\\
   e_\dd[n] &= \mat{C}_{1}\vc{\xi}[n] - \bar{u}_\dd[n],~
   v_\dd[n] = \mat{C}_{2}\vc{\xi}[n],\\
   \bar{u}_\dd[n] &= (k \ast v_\dd)[n],
  \end{split}
  \label{eq:lifted_sys}
 \end{equation}
 where $\ast$ stands for convolution, and the pertinent operators 
 $\mat{A}_\dd$, $\B$, 
 $\mat{C}_{1}$ and $\mat{C}_{2}$ are given as follows: 
 First, 
 $\B$ is a linear (infinite-dimensional) operator defined by
 \begin{equation}
  \begin{split}
   \B: L^2[0,T] &\rightarrow \Real^{\nu+1},\\
   \tw \mapsto \B\tw 
   &=
    \begin{bmatrix}
     \B_1\tw\\
     \B_2\tw
    \end{bmatrix}\\
   &=
    \begin{bmatrix}
      \displaystyle
      \int_{0}^{T}\ee^{\mat{A}(T-\tau)}\mat{B}\tw(\tau)\dd\tau\\
      \displaystyle
      \int_{0}^{T-d}\mat{C}\ee^{\mat{A}(T-d-\tau)}\mat{B}\tw(\tau)\dd\tau\\
    \end{bmatrix}
  \end{split}
  \label{eq:B}
 \end{equation}

 The matrices $\mat{A}_\dd$, $\mat{C}_1$, and $\mat{C}_2$ 
 in \eqref{eq:lifted_sys} are
 defined by
 \[
  \begin{split}
   \mat{A}_\dd 
   &:= \begin{bmatrix}
	\ee^{\mat{A}T}&0&0\\\
        \mat{C}\ee^{\mat{A}(T-d)}&0&0\\0&\mat{B}_m&\mat{A}_m
    \end{bmatrix}
   \in\Real^{(\nu+1+m)\times(\nu+1+m)},\\
   \mat{C}_{1} &:= [0, 0, \mat{C}_m]\in\Real^{1\times(\nu+1+m)},\\
   \mat{C}_{2} &:= [\mat{C}, 0, 0]\in\Real^{1\times(\nu+1+m)},
  \end{split}
 \]
 where $\mat{A}_m$, $\mat{B}_m$, and $\mat{C}_m$ are state-space realization matrices
 of the discrete-time delay
 $z^{-m}$.
\end{prop}
\begin{IEEEproof}
 See Appendix \ref{app:proof1}.
\end{IEEEproof}

\begin{figure}[t]
 \centering
 \includegraphics[width=0.7\linewidth]{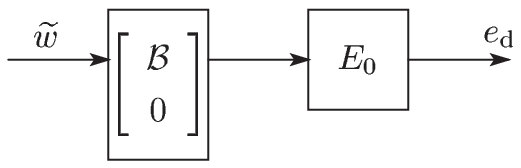}
 \caption{Factorization of $\tE$.}
 \label{fig:theorem1}
\end{figure}

The state-space representation \eqref{eq:lifted_sys}
then gives the transfer function of the lifted system $\tE$ as
\begin{equation}
 \hat{\tE}(z) = \hat{\G}_{1}(z) - \hat{K}(z)\hat{\G}_{2}(z),
 \label{eq:Ez}
\end{equation}
where
\[
  \hat{\G}_{i}(z) := \mat{C}_{i}(z\mat{I}-\mat{A}_\dd)^{-1}
   \left[\begin{array}{c}\B\\0\end{array}\right],\quad i=1,2.
\]
Put
\begin{equation}
 \hat{E}_0(z) := \bigl(\mat{C}_{1}-\hat{K}(z)\mat{C}_{2}\bigr)(z\mat{I}-\mat{A}_\dd)^{-1}.
 \label{eq:E0}
\end{equation}
Note that $E_0$ is a finite-dimensional discrete-time system.
Then the lifted system $\tE(z)$ in \eqref{eq:Ez} can be 
factorized (see Fig.~\ref{fig:theorem1}) as
\begin{equation}
 \hat{\tE}(z) = \hat{E}_0(z)\left[\begin{array}{c}\B\\0\end{array}\right].
 \label{eq:tE}
\end{equation}

\subsection{Norm-equivalent finite-dimensional system} 
\label{subsec:norm-equivalent}

The lifted system $\tE$ given in \eqref{eq:lifted_sys}, 
or its transfer function $\hat{\tE}$
in \eqref{eq:Ez},
involves an infinite-dimensional operator 
$\B: L^2[0,T) \rightarrow \Real^{\nu+1}$.
Introducing the dual operator~\cite{Yam93}
$\B^\ast: \Real^{\nu+1} \rightarrow L^2[0,T)$ of $\B$,
and composing this with $\B$, 
we can obtain a norm-equivalent finite dimensional system of 
the infinite-dimensional system $\tE$.

The dual operator $\B^\ast$ of $\B$ is given by~\cite{Yam93}
\begin{equation*}
 \begin{split}
  \B^\ast &= \left[\begin{array}{cc}\B_1^\ast&\B_2^\ast\end{array}\right],\quad
  \B_1^\ast(\theta) := \mat{B}^\top \ee^{\mat{A}^\top (T-\theta)},\\
  \B_2^\ast(\theta) &:= \ind_{[0,T-d)}(\theta)\mat{B}^\top 
    \ee^{\mat{A}^\top (T-d-\theta)}\mat{C}^\top,
  \quad \theta \in [0,T),
 \end{split}
\end{equation*}
where $\ind_{[0,T-d)}$ is the {characteristic} function of 
the interval $[0,T-d)$, that is,
\[
 \ind_{[0,T-d)}(\theta) := \begin{cases}
	1,\qquad \theta \in [0,T-d), \\
	0,\qquad \text{otherwise}.
 \end{cases}
\]
Then we have the following lemma:
\begin{lemma}\label{lemma:BB}
 The operator $\B\B^\ast$ is a positive semi-definite matrix given by
 \begin{equation}
  \begin{split}
   \B\B^\ast &=
   \begin{bmatrix}
   \B_1\B_1^\ast & \B_1\B_2^\ast\\
   \B_2\B_1^\ast & \B_2\B_2^\ast
   \end{bmatrix}\\
   &=
   \begin{bmatrix}
   \mat{M}(T)& \ee^{\mat{A}d}\mat{M}(T-d)\mat{C}^\top \\ 
   \mat{CM}(T-d)\ee^{\mat{A}^\top d}&\mat{CM}(T-d)\mat{C}^\top
   \end{bmatrix}
  \end{split}
  \label{eq:chol}
 \end{equation}
 where $\mat{M}(\cdot)$ is defined by
 \[
  \mat{M}(t):=\int_0^t \ee^{\mat{A}\theta}\mat{BB}^\top \ee^{\mat{A}^\top \theta}\dd\theta
   \in\Real^{\nu\times\nu},\quad t\geq 0.
 \]
\end{lemma}
\begin{IEEEproof}
We first prove $\B_1\B_2^\ast=\ee^{\mat{A}d}\mat{M}(T-d)\mat{C}^\top $.

For every $u\in\Real$, we have
\begin{equation*}
 \begin{split}
  &\B_1\B_2^\ast u
  \quad = \int_0^{T}  \ee^{\mat{A}(T-\theta)}\mat{B}\bigl(\B_2^\ast(\theta) u\bigr)\dd\theta\\
  &\quad = \int_0^{T}  \!\! \ee^{\mat{A}(T-\theta)}\mat{B}
    \left(\ind_{[0,T-d)}(\theta)\mat{B}^\top 
    \ee^{\mat{A}^\top (T-d-\theta)}\mat{C}^\top  u\right)\dd\theta\\
  &\quad = \ee^{\mat{A}d}\int_0^{T-d}\ee^{\mat{A}(T-d-\theta)}\mat{BB}^\top 
    \ee^{\mat{A}^\top (T-d-\theta)}\dd\theta\; \mat{C}^\top  u\\
  &\quad = \ee^{\mat{A}d}\mat{M}(T-d)\mat{C}^\top  u.
 \end{split}
\end{equation*}
Similarly, we can prove the equalities $\B_1\B_1^\ast=\mat{M}(T)$ 
and $\B_2\B_2^\ast=\mat{CM}(T-d)\mat{C}^\top$.
\end{IEEEproof}

\begin{remark}
The matrix 
$\mat{M}(t)$ 
can be computed via {the} matrix exponential {formula}~\cite{Loa78}:
\begin{equation*}
 \begin{split}
  \mat{M}(t)&=\mat{F}_{22}^\top (t)\mat{F}_{12}(t),\\
  \left[\begin{array}{cc}
   \mat{F}_{11}(t)&\mat{F}_{12}(t)\\0&\mat{F}_{22}(t)\end{array}\right]
  &:=\exp\left\{\left[\begin{array}{cc}-\mat{A}&\mat{BB}^\top \\0&\mat{A}^\top 
  \end{array}\right]t\right\}.
 \end{split}
\end{equation*}
By this formula, we can easily compute the matrices 
$\mat{M}(T)$ and $\mat{M}(T-d)$ in \eqref{eq:chol}
without performing a numerical integration.
\end{remark}

\begin{figure}[t]
 \centering
 \includegraphics[width=0.7\linewidth]{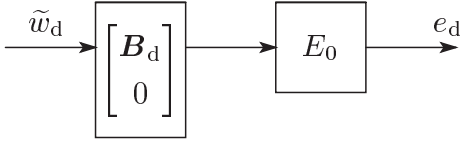}
 \caption{Discrete-time system  $\Td$.}
 \label{fig:theorem1_2}
\end{figure}

From Lemma \ref{lemma:BB},
$\B\B^\ast$ is a positive semi-definite matrix and hence
there exists a matrix $\mat{B}_\dd$ such that 
$\B\B^\ast=\mat{B}_\dd \mat{B}_\dd^\top $.
With matrix $\mat{B}_\dd$ and discrete-time system $E_0$
given in \eqref{eq:E0}, 
define a finite-dimensional discrete-time system by
\[
 \Td:=E_0\left[\begin{array}{c}\mat{B}_\dd\\0\end{array}\right].
\]
See Fig.~\ref{fig:theorem1_2} for the block diagram of $\Td$.
Then the
discrete-time system $\Td$
is equivalent to the original sampled-data error system $\E$
in Fig.~\ref{fig:error_sys} with respect to 
their $H^\infty$ norm
as described in the following theorem:
\begin{theorem}
\label{thm:norm}
Assume that the sampled-data error system $\E$ gives an
operator belonging to $\Bd(L^2,\ell^2)$, 
the set of all bounded linear operators of $L^2$ into $\ell^2$.
Then the discrete-time system $\Td$ belongs to 
$\Bd(\ell^2,\ell^2)$ and
equivalent to $\E$ with respect to their $H^\infty$ norm,
that is, 
$\|\E\|_\infty = \|\Td\|_\infty$.
\end{theorem}
\begin{IEEEproof}
First, the {equality} in \eqref{eq:EtotE}
and $\E\in\Bd(L^2,\ell^2)$ give
$\|\E\|_\infty = \|\tE\|_\infty < \infty$.
Using the factorization \eqref{eq:tE}, we have
\begin{equation*}
 \begin{split}
  \|\E\|_\infty^2 = \|\tE\|_\infty^2
  &=\left\|E_0\left[\begin{array}{c}
    \B\\0\end{array}\right]\right\|_\infty^2\\
  &= \left\|E_0
	\left[\begin{array}{c}\B\\0\end{array}\right]
	\left[\begin{array}{c}\B\\0\end{array}\right]^\ast E_0^\ast
  \right\|_\infty\\
  &= \left\|E_0
	\left[\begin{array}{c}\mat{B}_\dd\\0\end{array}\right]
	\left[\begin{array}{c}\mat{B}_\dd\\0\end{array}\right]^\top E_0^\ast
  \right\|_\infty
  = \left\|\Td\right\|_\infty^2.
 \end{split}
\end{equation*}
\end{IEEEproof}

Thus the sampled-data $H^\infty$ optimization (Problem \ref{prob:design}) 
is equivalently transformed to discrete-time $H^\infty$ optimization.
A MATLAB code for the $H^\infty$-optimal fractional delay filter is available
on the web at \cite{WEB-FDF}.
Moreover, if we assume that the filter $K(z)$ is an FIR filter,
the design is reduced to a convex optimization with a linear matrix inequality.
See \cite{NagYam05,Nag11} for details.

\subsection{Closed-form solution under a first-order assumption}
\label{subsec:closed-form}

Assume that the weighting function $\hat{W}(s)$ is 
a first-order low-pass filter
with cutoff frequency $\omega_\cc>0$:
\begin{equation}
 \hat{W}(s) = \frac{\omega_\cc}{s+\omega_\cc}.
 \label{eq:Fs}
\end{equation}
Under this assumption, a closed-form solution for
the optimal filter is obtained \cite{NagYam03,NagYam05}:
\begin{theorem}
\label{thm:closed-form}
Assume that $\hat{W}(s)$ is given by \eqref{eq:Fs}. Then
the optimal filter $\hat{K}(z)$ is given by
\begin{equation}
 \hat{K}(z) = a_0(d)z^{-m} + a_1(d) z^{-m-1},
 \label{eq:opt-filter}
\end{equation}
where
\[
 \begin{split}
  a_0(d) := \frac{\sinh\left(\omega_\cc(T-d)\right)}
    {\sinh(\omega_\cc T)},~
  a_1(d) := \ee^{-\omega_\cc T}\bigl(\ee^{\omega_\cc d} - a_0(d)\bigr).
 \end{split}
\]
Moreover, the optimal value of $\|\E\|_\infty$ is given by
\begin{equation}
 \|\E\|_\infty 
  = \sqrt{\frac{\omega_\cc\sinh(\omega_\cc d)\sinh(\omega_\cc (T-d))}
    {\sinh(\omega_\cc  T)}}.
 \label{eq:optimal}
\end{equation}
\end{theorem}

Since the optimal filter $\hat{K}(z)$ in \eqref{eq:opt-filter}
is a function of the fractional delay $d$ and the integer delay $m$,
the filter can be used as a \emph{variable fractional delay filter}~\cite{LVKL96}.
\begin{remark}
Fix $d>0$ and $m\in\Z_+$ arbitrarily. 
By definition, we have $T-d<T$.
It follows that as $\omega_\cc \rightarrow \infty$, we have
$a_0(d)\rightarrow 0$, $a_1(d)\rightarrow 0$,
and
$\|E\|_\infty \rightarrow \infty$.
This means that if the original analog signals contain higher frequency
components (far beyond the Nyquist frequency), 
the worst-case input signal becomes more severe, and
the $H^\infty$-optimal filter becomes closer to $0$.
\end{remark}

\section{Design Examples}
\label{sec:example}

We here present design examples of fractional delay filters.

The design parameters are as follows:
the sampling period $T=1$ (sec), 
the delay $D=5.5$ (sec), that is, $m=5$ and $d=0.5$.
The frequency-domain characteristic of analog signals
to be sampled is modeled by
\begin{equation*}
 \hat{W}(s)=\frac{\omega_\cc}{s+\omega_\cc}, \quad \omega_\cc = 0.1.
\end{equation*}
Note that $\hat{W}(s)$ has the cutoff frequency $\omega_\cc =0.1$ (rad/sec)
$\approx 0.016$ (Hz),
which is below the Nyquist frequency $\pi$ (rad/sec) $=0.5$ (Hz).

\begin{figure}[t]
 \centering
 \includegraphics[width=\linewidth]{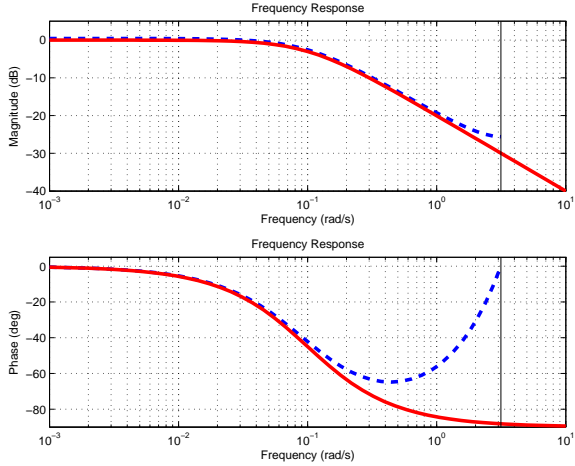}
 \caption{
  Bode plot of $\hat{W}(s)$ (solid) and its impulse-invariant discretization
  $\hat{W}_\dd(z)$ (dash). The vertical line shows the Nyquist frequency.
 }
 \label{fig:weights}
\end{figure}

We compare the sampled-data $H^\infty$ optimal filter obtained 
by Theorem \ref{thm:closed-form}
with conventional FIR filters designed by
discrete-time $H^2$ optimization~\cite{LVKL96},
which minimizes the cost function \eqref{eq:H2norm}.
The weighting function $\hat{W}_\dd(z)$ in \eqref{eq:E2} or \eqref{eq:H2norm}
is chosen as the impulse-invariant discretization~\cite{Jac00} of $\hat{W}(s)$.
Fig.~\ref{fig:weights} shows the Bode plots of 
$\hat{W}(s)$ and $\hat{W}_\dd(z)$.

\begin{figure}[t]
 \centering
 \includegraphics[width=\linewidth]{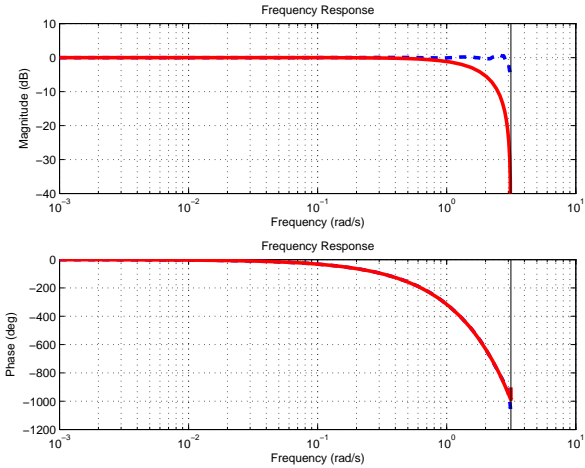}
 \caption{
  Bode plot of filters: sampled-data $H^\infty$ design (solid),
  and discrete-time $H^2$ design (dash).
 }
 \label{fig:filter}
\end{figure}

The transfer function of the proposed filter is given by
\[
 \hat{K}(z) = z^{-5}\left(0.4994+0.4994z^{-1}\right).
\]
Fig.~\ref{fig:filter} shows the Bode plots
of the designed filters.
As illustrated in Fig.~\ref{fig:filter}, the $H^2$ optimal filter
is closer to the ideal filter \eqref{eq:ideal_FD_fourier}
as expected, so that it appears better in the context of 
the conventional design methodology. 

\begin{figure}[t]
 \centering
 \includegraphics[width=\linewidth]{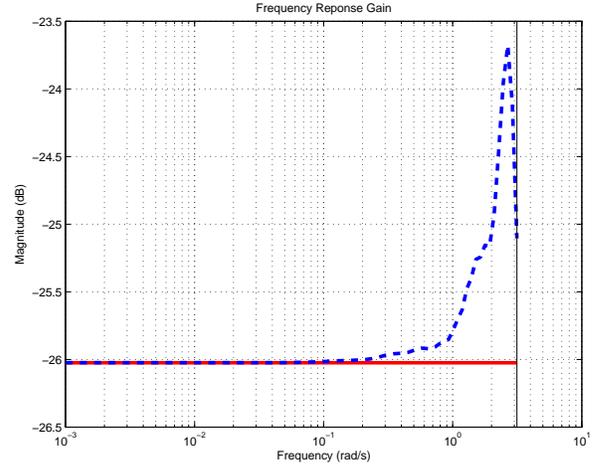}
 \caption{
  Frequency response gain of error system $\E$ in Fig.~\ref{fig:error_sys}:
  sampled-data $H^\infty$ design (solid),
  and discrete-time $H^2$ design (dash).
 }
 \label{fig:Tew}
\end{figure}

However, the $H^2$ optimal filter exhibits {\em much larger errors}
in the high-frequency domain as shown in
Fig.~\ref{fig:Tew} that shows the frequency response gain of 
the sampled-data error system $\E$ shown in Fig.~\ref{fig:error_sys}.
%
This is because the conventional designs cannot take into account
the frequency response of the source analog signals 
while the present method does.

\begin{figure}[t]
 \centering
 \includegraphics[width=\linewidth]{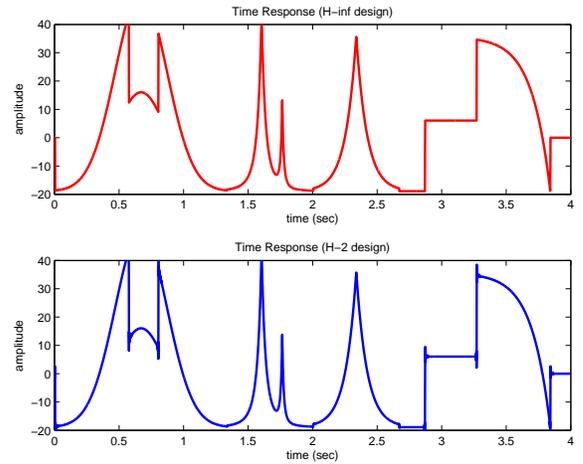}
 \caption{
  Time response: sampled-data $H^\infty$ design (above) 
  and discrete-time $H^2$ design (below)
  with sampling frequency $1000$ (Hz).}
 \label{fig:time_resp}
\end{figure}

\begin{figure}[t]
 \centering
 \includegraphics[width=\linewidth]{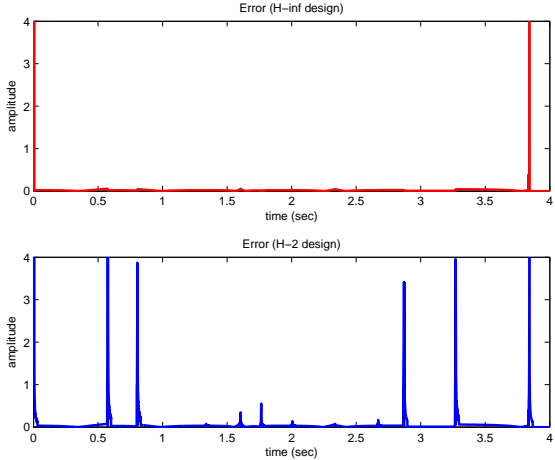}
 \caption{
  Absolute value of reconstruction error: sampled-data $H^\infty$ design (above) 
  and discrete-time $H^2$ design (below).}
 \label{fig:error}
\end{figure}

To see the difference between the present filter and the conventional one, 
we show the time response against a piecewise regular signal
produced by the \texttt{MakeSignal} function of WaveLab \cite{WaveLab}
in Fig.~\ref{fig:time_resp}.
The present method is superior
to the conventional one that shows much ringing at edges
of the wave. 
To see the difference more finely, we show the reconstruction
error in Fig.~\ref{fig:error}.
The $H^2$-optimal filter has much larger errors around edges of the signal
than the proposed $H^\infty$-optimal one.
In fact, the $L^2$ norm of the error is
$1.34\times 10^{-2}$ for $H^\infty$ design and $2.07\times 10^{-2}$
for $H^2$ design.
This illustrates the effectiveness of our method.


\section{Conclusion}
\label{sec:conc}

We have presented a new method of designing  fractional delay
filters via sampled-data $H^\infty$ optimization.
An advantage here is that an {\em optimal analog 
performance\/} can be attained.  
The optimal design problem can be equivalently transformed to
discrete-time $H^\infty$ optimization,
which is easily executed by standard numerical optimization
toolboxes.
A closed-form solution is given 
when the frequency distribution of the input analog 
signal is modeled as a first-order low-pass filter.
Design examples show that 
the $H^\infty$-optimal filter exhibits a much more satisfactory performance
than {the} conventional $H^2$-optimal filter.

\section*{Acknowledgement}
This research is supported in part by the JSPS Grant-in-Aid for Scientific Research (B) No.~24360163,
(C) No.~24560543, and Grant-in-Aid for Exploratory Research No.~22656095.

\appendices

\section{Proof of Proposition \ref{prop:ss}}
\label{app:proof1}

From the relation \eqref{eq:E},
the lifted system $\tE$ is described as
(see also Fig.~\ref{fig:lifted_sys})
\[
 \begin{split}
  \tE 
    &= \samp{T}\ee^{-Ds}W\ilift - K\samp{T}W\ilift\\
    &= \samp{T}\ee^{-mTs}\ee^{-ds}W\ilift-K\samp{T}W\ilift\\
    &= z^{-m}\samp{T}\ee^{-ds}W\ilift - K\samp{T}W\ilift\\
    &= z^{-m}y_\dd -Kv_\dd,
 \end{split}
\]
where $y_\dd:=\samp{T}\ee^{-ds}W\ilift\tw$ and $v_\dd:=\samp{T}W\ilift\tw$.
From the state-space representation of $W$ in \eqref{eq:Fss},
for any $t_1$ and $t_2$ such that $0\leq t_1 \leq t_2 < \infty$, we have
\[
 \vc{x}(t_2) = \ee^{\mat{A}(t_2-t_1)}\vc{x}(t_1) 
  + \int_{t_1}^{t_2}\ee^{\mat{A}(t_2-t)}\mat{B}w(t)\dd t.
\]
Putting $t_1:=nT$ and $t_2:=(n+1)T$ for $n\in\Z_{+}$ gives
\[
 \vc{x}(nT+T) = \ee^{\mat{A}T}\vc{x}(nT) 
  + \int_0^T \ee^{\mat{A}(T-\tau)}\mat{B}w(nT+\tau)\dd\tau.
\]
Define $\vc{x}_1[n]:=\vc{x}(nT)$ and $\tw[n]:=(\lift w)[n]$.
Then we have
\begin{equation}
 \vc{x}_1[n+1] = \ee^{\mat{A}T}\vc{x}_1[n] + \B_1\tw[n],
 \label{eq:app1-1}
\end{equation}
where $\B_1$ is defined in \eqref{eq:B}.
On the other hand, from \eqref{eq:Fss}, we have
$v(t) = \mat{C}\vc{x}(t)$ for $t\in\Real_+$.
Putting $t_1:=nT$ and $t_2:=nT+\theta$ for $n\in\Z_{+}$ 
and $\theta\in[0,T)$, we have
\begin{equation}
 \begin{split}
  v(nT+\theta) &= \mat{C}\vc{x}(nT+\theta)\\
    &= \mat{C}\ee^{\mat{A}\theta}\vc{x}(nT) 
    + \int_0^\theta \mat{C}\ee^{\mat{A}(\theta-\tau)}
       \mat{B}w(nT+\tau)\dd\tau.
 \end{split}
 \label{eq:lift_v}
\end{equation}
By this, we have
\begin{equation}
 v_\dd[n] = v(nT) = \mat{C}\vc{x}_1[n].
 \label{eq:app1-2}
\end{equation}
Next, from \eqref{eq:lift_v}, we have
\[
 \begin{split}
  y_\dd[n] &= v(nT-d)
   = v(nT-T+T-d)\\
   &= \mat{C}\ee^{\mat{A}(T-d)}\vc{x}_1[n-1]\\
   &\qquad + \int_0^{T-d}\mat{C}\ee^{\mat{A}(\theta-\tau)}
     \mat{B}\tw[n-1](\tau)\dd\tau.
 \end{split}
\]
Put $x_2[n]:=y_\dd[n]$. Then we have
\begin{equation}
 \begin{split}
  x_2[n+1] &= \mat{C}\ee^{\mat{A}(T-d)}\vc{x}_1[n] + \B_2\tw[n],\\
  y_\dd[n] &= x_2[n],
 \end{split}
 \label{eq:app1-3}
\end{equation}
where $\B_2$ is defined in \eqref{eq:B}.
By the relation 
\[
 u_\dd[n]=y_\dd[n-m]=z^{-m}y_\dd[n],
\]
and the state-space matrices $\mat{A}_m$, $\mat{B}_m$, 
and $\mat{C}_m$ 
for $m$-step delay $z^{-m}$,
we have
\begin{equation}
 \begin{split}
  \vc{x}_3[n+1] &= \mat{A}_m \vc{x}_3[n] + \mat{B}_m y_\dd[n],\\
  u_\dd[n] &= \mat{C}_m \vc{x}_3[n].
 \end{split}
 \label{eq:app1-4}
\end{equation}
Combining \eqref{eq:app1-1}, \eqref{eq:app1-2}, \eqref{eq:app1-3},
and \eqref{eq:app1-4} all together
gives the state-space representation \eqref{eq:lifted_sys} with
$\vc{\xi}^\top := [\vc{x}_1^\top, x_2, \vc{x}_3^\top]^\top$.
\hfill \IEEEQED

\ifCLASSOPTIONcaptionsoff
  \newpage
\fi



%


\end{document}